\documentclass[conference]{IEEEtran}

\usepackage{cite}
\usepackage{amsmath,amssymb,amsfonts}
\usepackage{graphicx}
\usepackage{algorithmic}
\usepackage{textcomp}
\usepackage{xcolor}
\usepackage{stfloats}
\usepackage[acronym]{glossaries}
\usepackage{comment}

\usepackage{etoolbox}

\usepackage{geometry}
\newtoggle{submission}
\toggletrue{submission}

\begin{document}

\title{Learning the LoS Skyline from LEO Satellite Observations for Proactive Handover}

\author{\IEEEauthorblockN{Marius Corici\IEEEauthorrefmark{1}\IEEEauthorrefmark{2}, Manar Zaboub\IEEEauthorrefmark{1}, Fabian Eichhorn\IEEEauthorrefmark{1}, Hauke Buhr\IEEEauthorrefmark{1}}
\IEEEauthorblockA{\IEEEauthorrefmark{1}Fraunhofer FOKUS, Berlin, Germany}
\IEEEauthorblockA{\IEEEauthorrefmark{2}Technische Universit\"at Berlin, Germany\\
Email: \{marius-iulian.corici, manar.zaboub, fabian.eichhorn, hauke.buhr\}@fokus.fraunhofer.de}
}

\maketitle

\begin{abstract}

In non-terrestrial network deployments, local obstructions may block line-of-sight satellite links before the satellite reaches the geometric elevation mask, causing abrupt and unplanned handovers. To address this limitation, this paper proposes a map-free method for learning the local LoS skyline, defined as the obstruction elevation over azimuth, from binary availability labels derived from passive satellite signal observations at the terminal. The problem is formulated as a binary classification task in the azimuth-elevation space, where the skyline is extracted as the decision boundary of the learned blockage probability surface. Two complementary estimators are investigated, namely a Gaussian Process (GP) classifier and a neural multilayer perceptron (MLP) with circular azimuth encoding and Monte Carlo Dropout uncertainty indicators. The learned obstruction surface is then combined with satellite ephemeris information through EphemerisWindow, a trajectory-level prediction method that estimates future LoS termination events before the serving link is lost. The results show that both learned estimators improve the skyline reconstruction compared with empirical bracketing and enable proactive handover preparation without requiring 3D building maps, sky cameras, or additional environmental sensing.
\end{abstract}

\begin{IEEEkeywords}
5G NTN, Mega-Constellation, NTN Handover
\end{IEEEkeywords}

\section{Introduction}

Low Earth Orbit (LEO) satellite systems are becoming an integral component of non-terrestrial network (NTN) architectures, as they extend the reach of mobile networks beyond the footprint of terrestrial infrastructure~\cite{3gpp_tr38821}. However, their inherent mobility also changes the nature of the handover problem. A serving satellite remains visible only for a limited time, and the handover decision must therefore consider the current radio condition together with the future availability of the satellite link.

\begin{figure}[!b]
    \centering
    \includegraphics[width=0.9\columnwidth]{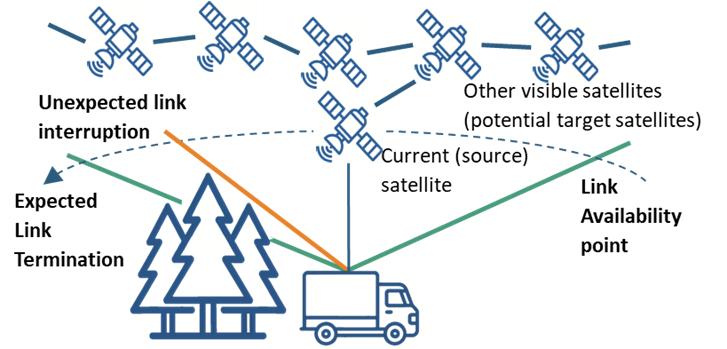}
    \caption{Local LoS loss above the elevation mask}
    \label{fig:use_case}
\end{figure}

In open-sky conditions, this future availability can be derived from satellite ephemeris information and a geometric elevation mask. In obstruction-rich environments, commonly encountered on the earth's surface, this assumption no longer holds. Local objects such as buildings, terrain, vegetation, or roadside structures may block the Line-of-Sight (LoS) path while the satellite is still above the elevation threshold. As illustrated in Fig.~\ref{fig:use_case}, the terminal may observe several satellites that are geometrically visible, while only a subset of them remains usable from the local position. Without knowledge of this local obstruction profile, the handover process can only react after the serving link has significantly degraded or has already been lost~\cite{corici2025servicecontinuity,corici2026reactive}.

To address this limitation, this paper proposes to learn the local LoS skyline directly from passive satellite signal observations. The LoS skyline is defined as the obstruction elevation over azimuth and is inferred from binary blocked and unblocked observations collected during normal terminal operation. Given the current satellite trajectory, the terminal can then estimate where this trajectory intersects the learned local horizon line, and therefore how much usable LoS time remains before the current satellite disappears. This remaining time provides the basis for triggering the handover preparation phase before the reference signal is lost.

The contribution of the paper is threefold. First, the LoS skyline learning problem is formulated as a binary classification task in the azimuth-elevation space. Second, two estimators, namely a Gaussian Process (GP) classifier and a neural multilayer perceptron (MLP), are investigated for reconstructing the blockage probability surface. Third, the learned surface is combined with satellite ephemeris information through EphemerisWindow, enabling trajectory-level prediction of future LoS termination events as an advance-warning input for handover preparation.

The considered setting is a fixed terminal whose geographic position and satellite ephemerides are known with sufficient accuracy to compute satellite azimuth and elevation. The method is therefore not GNSS-free. The local obstruction profile is assumed to remain stable during the learning interval. The resulting skyline is specific to the terminal location and must be updated or relearned after relocation or a relevant environmental change. The intended use case is therefore a fixed NTN terminal rather than a continuously moving UE. As illustrated in Fig.~\ref{fig:use_case}, geometric visibility above the elevation mask does not necessarily imply local LoS availability at the terminal position.

\section{Related Work}\label{sec:related}

LEO satellite handover has mainly been addressed from the perspective of satellite selection, access delay, collision probability, and proactive scheduling~\cite{park2021leo,lee2023handover,he2026preho,jang2026proactive}. These mechanisms are essential for selecting the target satellite and preparing the control-plane procedure. However, the visibility information used by such mechanisms is commonly derived from satellite trajectory geometry. This is sufficient when the elevation mask is the dominant visibility constraint, but it becomes incomplete when the relevant event is the local disappearance of the reference signal before the satellite reaches the geometric mask.

Obstruction awareness has also been studied through GNSS shadow matching, skymask estimation, satellite visibility prediction, and camera-based sky segmentation~\cite{zheng2024gnss,hornillo2020prediction,xu2020machine,li2023machine,lee2020skymask,wang2024skygvio}. These approaches underline the importance of the local sky visibility boundary. However, they typically depend on three-dimensional city models, optical sensors, or signal-level classifiers designed for individual observations. As such, they do not directly provide a map-free obstruction surface that can be queried for future LEO satellite trajectories.

The approach considered in this paper provides the missing local visibility layer between satellite trajectory prediction and handover preparation.

\section{System Model}\label{sec:sysmod}

A fixed terminal is considered in a LEO-based NTN deployment. The terminal is assumed to know its geographic position and the satellite ephemerides, allowing it to compute the future azimuth and elevation of each satellite. It also receives periodic radio measurement opportunities, such as channel state information reference signal (CSI-RS) or synchronization signal block (SSB) measurements in 5G New Radio (NR). These measurements provide an operational indication of whether the expected satellite reference signal remains available from the corresponding direction.

A geometric elevation mask $\theta_{\min}$ defines the minimum elevation above which a satellite is considered geometrically visible. However, geometric visibility does not imply local LoS availability.
A satellite may remain above the elevation mask while its reference signal becomes unavailable because the LoS path is blocked by the local environment. In a practical implementation, a blocked label can be declared when the expected reference signal is not detected or remains below a configured RSRP or SINR threshold for consecutive measurement occasions. Requiring consecutive observations reduces isolated labels caused by short fades. Nevertheless, reflections, attenuation, rain, and temporary obstructions may still create imperfect LoS/NLoS labels. For observation $i$, associated with one of the observed satellites, the terminal records

\begin{equation}
    o_i = (\phi_i,\theta_i,b_i),
\end{equation}
where $\phi_i \in [0^\circ,360^\circ)$ denotes the satellite azimuth obtained from the ephemeris, $\theta_i \in [\theta_{\min},90^\circ]$ denotes the corresponding elevation, and $b_i \in \{0,1\}$ is the binary availability label, with $b_i=1$ denoting a blocked or unavailable measurement. The index $i$ identifies an observation sample collected across the set of observed satellites.

\begin{equation}
    \mathcal{D} = \{(\phi_i,\theta_i,b_i)\}_{i=1}^{N}.
\end{equation}

This buffer is used to infer the local obstruction boundary at which a satellite disappears from LoS, without relying on a prior map, a sky camera, or an additional environmental sensor.

\section{LoS Skyline Formulation}\label{sec:losskyline}

The local LoS skyline is defined as a function
\begin{equation}
    s^\star : [0^\circ,360^\circ) \rightarrow [\theta_{\min},90^\circ],
\end{equation}
where $s^\star(\phi)$ gives the obstruction elevation in the azimuth direction $\phi$. A satellite located at $(\phi,\theta)$ remains locally available when its elevation is above this skyline and disappears from LoS when its trajectory crosses the skyline from above. The LoS condition is therefore given by
\begin{equation}
    \theta > s^\star(\phi).
\end{equation}
The true skyline is unknown to the terminal and must be inferred from the training data set $\mathcal{D}$.

To achieve this, LoS skyline learning is formulated as binary classification in the azimuth-elevation space. The classifier estimates the blockage probability
\begin{equation}
    P(b=1 \mid \phi,\theta),
\end{equation}
and the estimated skyline is extracted as the decision boundary of this probability surface. For a decision threshold $\tau$, the estimated skyline is written as
\begin{equation}
    \hat{s}(\phi;\tau) = \{\theta : P(b=1 \mid \phi,\theta)=\tau\}.
\end{equation}
The central contour $\tau=0.5$ represents the maximum-likelihood obstruction boundary, while lower thresholds define conservative contours that can be used to anticipate the LoS disappearance and start handover preparation earlier.

Since azimuth is circular, a scalar representation would introduce an artificial discontinuity between $0^\circ$ and $360^\circ$. Therefore, each observation is encoded as
\begin{equation}
    x_i =
    \left(
    \sin \phi_i,\,
    \cos \phi_i,\,
    \frac{\theta_i-\theta_{\min}}{\theta_{\max}-\theta_{\min}}
    \right),
\end{equation}
where $\theta_{\max}=90^\circ$. The elevation component is therefore min--max normalised to the interval $[0,1]$. This representation preserves angular continuity and provides a common input space for the estimators we will introduce in Section~\ref{sec:learning}.

\section{LoS Skyline Learning Methods}\label{sec:learning}

The formulation introduced in Section~\ref{sec:sysmod} requires an estimator for the blockage probability surface $P(b=1 \mid \phi,\theta)$. To capture different operating points, three methods are considered.  

\subsection{GP Classifier}
The GP classifier provides a probabilistic and data-efficient estimator for limited training data sets. It models the latent blockage function over the encoded input space by placing a GP prior on $f(x)$ and by mapping the resulting latent value to a blockage probability through a logistic link function~\cite{rasmussen2006gaussian}. Given the training data set $\mathcal{D}$, the model is trained on the encoded samples $x_i$ and their corresponding blockage indicators $b_i$, thereby learning a smooth probability surface from which the LoS skyline can be extracted.

The covariance function is defined as a radial basis function with constant amplitude,
\begin{equation}
    k(x,x') =
    C \exp\left(-\frac{\|x-x'\|^2}{2\ell^2}\right),
\end{equation}
where $C$ denotes the signal amplitude and $\ell$ denotes the characteristic length scale. The length scale determines the angular smoothness of the obstruction surface. Larger values produce broader and smoother contours, while smaller values allow sharper local transitions. The kernel hyperparameters are selected through marginal likelihood optimisation, allowing the GP to adapt the obstruction scale to the available observations.

\subsection{Neural MLP Classifier}
The neural multilayer perceptron provides a scalable estimator for larger training data sets and more irregular obstruction profiles. It approximates the same blockage probability surface by processing the encoded input $x_i$ through a fully connected network with smooth nonlinear activations. The resulting output logit is then mapped to a blockage probability through a sigmoid function, allowing the model to learn a flexible decision boundary from the encoded observations and to support skyline extraction through the same threshold-based procedure used for the GP estimator. The implemented MLP contains three hidden layers with 64 units each, using tanh activations and a dropout rate of 0.1. The model is trained for 400 epochs with a batch size of 256 using the Adam optimizer with a learning rate of $2\times10^{-3}$, by minimising the binary cross-entropy loss

\begin{equation}
    \mathcal{L}
    =
    - \sum_i
    \left[
    b_i \log \hat{p}_i
    +
    (1-b_i)\log(1-\hat{p}_i)
    \right],
\end{equation}
where $\hat{p}_i=P(b=1 \mid x_i)$ denotes the predicted blockage probability.

To obtain an uncertainty indicator, Monte Carlo Dropout is applied during inference~\cite{gal2016dropout}. For each azimuth direction, 30 stochastic forward passes are evaluated, and the standard deviation of the resulting skyline elevation estimates is used as the predictive uncertainty indicator. This provides a practical approximation of uncertainty while preserving the scalability of the neural estimator. The resulting standard deviations are used only as relative uncertainty indicators and are not interpreted as calibrated confidence values.

The MLP estimator is less constrained by the stationary RBF kernel used by the GP and can be trained on larger training data sets with approximately linear scaling in the number of samples. It is therefore suitable for long observation periods and irregular obstruction profiles. However, the neural decision boundary generally requires more data and more training iterations before it reaches the sharpness obtained by the GP in smooth scenarios.

\subsection{Empirical Bracketing Baseline}
The empirical bracketing method provides a model-free baseline based directly on blocked and unblocked observations. The azimuth range is divided into fixed angular bins of $10^\circ$. For each bin, the highest observed blocked elevation (i.e. the signal was not received when expected) and the lowest observed unblocked elevation are extracted. The skyline is then estimated as the midpoint between these two values,
\begin{equation}
    \hat{s}_{\mathrm{emp}}(\phi)
    =
    \frac{
    \theta^{\mathrm{blocked}}_{\max}(\phi)
    +
    \theta^{\mathrm{open}}_{\min}(\phi)
    }{2}.
\end{equation}

A wrapped Gaussian filter with a width of $1.2$ bins is applied across adjacent azimuth bins. Wider bins and stronger smoothing reduce variance in sparsely observed regions but may remove narrow obstruction features and bias sharp skyline transitions.

This baseline is transparent and requires no training. However, each bin is treated independently, and the method cannot infer reliable values in azimuth regions where few satellite passes have been observed. It is also sensitive to outliers near the obstruction boundary. As such, it provides a lower-complexity reference against which the GP and neural estimators can be evaluated.

\subsection{Skyline Extraction}

After training, the GP and MLP estimators are queried through the same extraction procedure. For each azimuth direction, the blockage probability is evaluated over the elevation interval $[\theta_{\min},\theta_{\max}]$. The skyline is obtained as the elevation at which the selected threshold $\tau$ is crossed. A monotonicity correction is applied along the elevation dimension before extraction, ensuring that the blockage probability does not increase with elevation. This produces a well-defined skyline contour for the learned estimators and allows the same threshold-based interpretation to be used for LoS termination prediction.

\iftoggle{submission}{
\section{EphemerisWindow for LoS Prediction}\label{sec:ephemwin}
}{
\section{EphemerisWindow for LoS Termination Prediction}\label{sec:ephemwin}
}

The learned blockage probability surface can be used to predict when a satellite that is still geometrically visible will disappear from LoS. Given the ephemeris of a satellite $s$, its future azimuth and elevation trajectory can be written as $(\phi_s(t),\theta_s(t))$. The learned estimator then provides the corresponding blockage probability along this trajectory. A LoS termination event is predicted when the trajectory crosses a selected blockage threshold $\tau_{\mathrm{pred}}$,
\begin{equation}
    t_s^\star =
    \min \left\{
    t > t_0 :
    P(b=1 \mid \phi_s(t),\theta_s(t)) > \tau_{\mathrm{pred}}
    \right\},
\end{equation}
where $t_0$ denotes the current time and $t_s^\star$ is the earliest predicted time at which satellite $s$ loses local LoS availability. The predicted advance warning is then given by
\begin{equation}
    d_s = t_s^\star - t_0 .
\end{equation}
If no threshold crossing is detected within the considered prediction horizon, the satellite is treated as locally available over that horizon.

The EphemerisWindow mechanism extends this prediction from the serving satellite to the set of satellites that are geometrically visible or expected to become geometrically visible within a future time window $H$. For each candidate satellite, the ephemeris is scanned over the interval $[t_0,t_0+H]$, and the learned blockage surface is queried along the corresponding trajectory. This provides one predicted LoS termination time per satellite trajectory, rather than a sequence of independent per-step blockage decisions.
This trajectory-level interpretation is essential for handover preparation. A per-step trigger may react to isolated probability crossings caused by uncertainty or sparse observations, thereby producing premature preparation events. EphemerisWindow instead evaluates whether the future trajectory of a satellite consistently intersects the learned obstruction surface. As such, the prediction is tied to the expected disappearance of the reference signal from LoS, which is the event that matters for handover preparation.

The output of EphemerisWindow is a ranked candidate list,
\begin{equation}
\mathcal{C}(t_0)
    =
    \mathrm{sort}_{d_s}^{\mathrm{desc}}
    \left(
    \{(s,d_s)\}_{s \in \mathcal{S}_H}
    \right),
\end{equation}
where $\mathcal{S}_H$ denotes the satellites considered within the prediction horizon and the candidates are sorted in descending order of their estimated remaining LoS time $d_s$. Satellites with longer predicted local availability can be preferred as handover targets, while satellites with imminent LoS termination can be avoided even if they are still geometrically visible. The mechanism therefore provides an obstruction-aware preparation signal, without evaluating a full handover execution procedure.
The prediction threshold $\tau_{\mathrm{pred}}$ is a design parameter. Lower values provide earlier warnings by using a conservative contour of the learned blockage surface, while higher values place the trigger closer to the central skyline estimate. The suitable value depends on the preparation time required by the network and on the desired trade-off between early warning and geometric tightness. In this paper, the threshold is used to evaluate the advance-warning capability of the learned skyline, rather than the complete performance of a mobility management procedure.

\section{Evaluation}\label{sec:eval}

\begin{figure*}[tb]
    \centering
    \includegraphics[width=\textwidth]{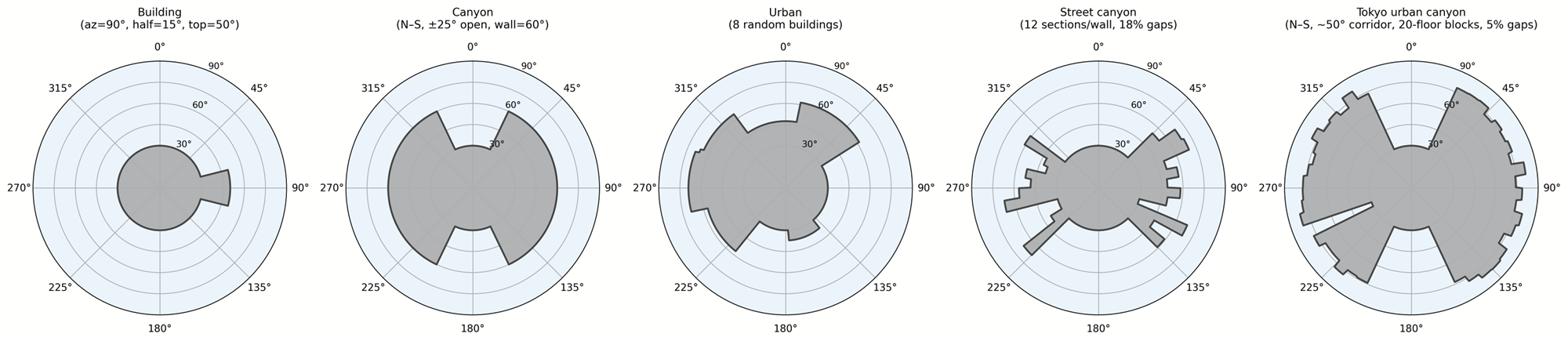}
    \caption{Evaluated synthetic LoS skyline geometries}
    \label{fig:geometries}
\end{figure*}

The evaluation considers a LEO constellation with $432$ satellites at $600$~km altitude, $65^\circ$ inclination, and $12$ orbital planes with $36$ satellites per plane. The terminal is located in Berlin, Germany, and applies a geometric elevation mask of $\theta_{\min}=30^\circ$. .Satellite positions are sampled at one-minute resolution over a two-orbit observation window, corresponding to approximately three hours and resulting in approximately $2105$ binary observations. Five synthetic obstruction geometries are considered, covering isolated, regular, and irregular LoS skyline profiles, as illustrated in Fig.~\ref{fig:geometries}. For the controlled reconstruction evaluation, the labels are generated directly from the known skyline according to whether $\theta_i \leq s^\star(\phi_i)$. No NR waveform, propagation, received-power, or reference-signal detection model is included. The reported results therefore isolate skyline reconstruction performance under ideal binary labels and do not constitute an end-to-end radio evaluation.

The empirical bracketing baseline, the Gaussian Process classifier, and the neural MLP classifier are evaluated using the same training data set. Skyline accuracy is measured through the root mean square error (RMSE), which quantifies the average angular deviation between the estimated skyline $\hat{s}(\phi)$ and the true skyline $s^\star(\phi)$ over $M$ uniformly sampled azimuth directions:
\begin{equation}
\mathrm{RMSE}
=
\sqrt{
\frac{1}{M}
\sum_{j=1}^{M}
\left(
\hat{s}(\phi_j)-s^\star(\phi_j)
\right)^2
}.
\end{equation}
Classification quality is measured through the false negative rate (FNR), the false positive rate (FPR), and the F1 score for the blocked class. LoS termination prediction is evaluated through the mean advance warning before the actual LoS disappearance event and the false alarm rate of the EphemerisWindow prediction.

\subsection{Skyline Reconstruction}

\begin{figure*}[tb]
    \centering
    \includegraphics[width=\textwidth]{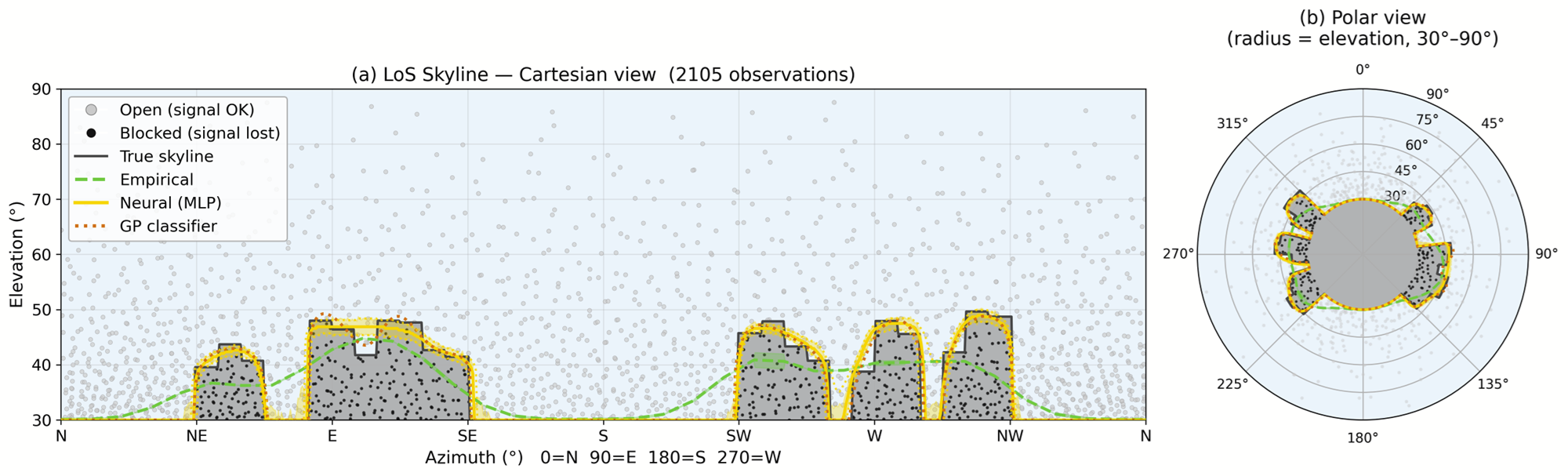}
    \caption{Representative LoS skyline reconstruction from 2105 satellite observations}
    \label{fig:probability_surface}
\end{figure*}

\begin{figure*}[tb]
    \centering
    \includegraphics[width=\textwidth]{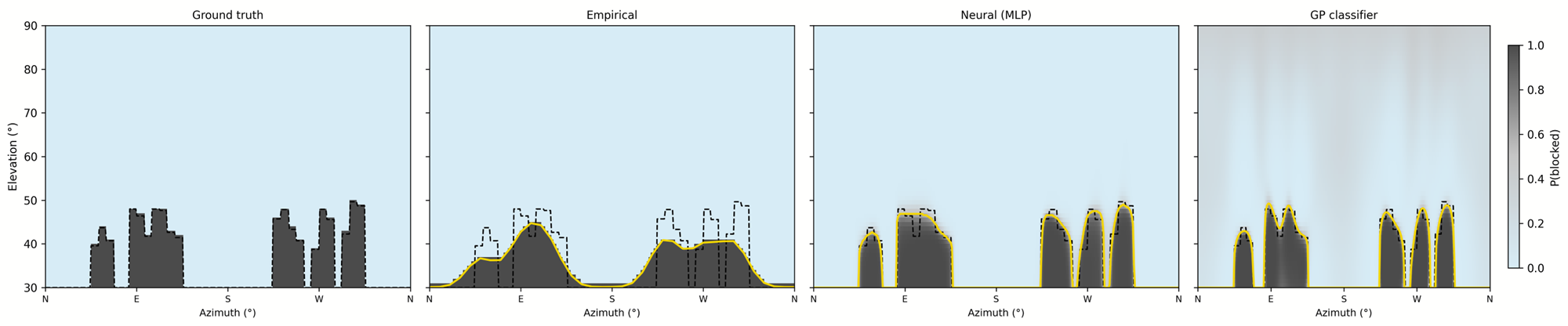}
    \caption{Ground-truth blockage region and estimator outputs with reconstructed skyline}
    \label{fig:skyline_reconstruction}
\end{figure*}

Fig.~\ref{fig:probability_surface} compares the reconstructed skylines with the binary observations. Fig.~\ref{fig:skyline_reconstruction} shows the estimator outputs, where the central contour defines the reconstructed skyline and the surrounding information indicates predictive uncertainty.

Table~\ref{tab:skyline_rmse} reports the mean RMSE over five runs. The runs vary the MLP initialisation and, for the randomised Urban profile, the obstruction geometry. The GP achieves the lowest error in four geometries and ties the MLP in the most irregular case, while empirical bracketing remains consistently less accurate.

\begin{table}[t]
\centering
\caption{Skyline RMSE in degrees after the two-orbit observation window.}
\label{tab:skyline_rmse}
\begin{tabular}{lccccc}
\hline
Estimator & Building & Canyon & Urban & Street & Tokyo \\
\hline
Empirical & 3.1 & 6.3 & 6.7 & 9.5 & 11.1 \\
GP & \textbf{0.6} & \textbf{0.9} & \textbf{2.3} & \textbf{2.9} & \textbf{5.0} \\
MLP & 1.9 & 2.2 & 3.0 & 4.6 & \textbf{5.0} \\
\hline
\end{tabular}
\end{table}

\subsection{Classification Accuracy}

Table~\ref{tab:classification} summarises the classification performance at the central threshold $\tau=0.5$. Both learned estimators substantially improve over the empirical baseline. The GP obtains the lowest missed-blockage rate and the highest F1 score, while the neural estimator reaches a comparable false positive rate. This confirms that the learned probability surfaces provide a more reliable representation of the local obstruction boundary than direct bracketing alone.

\begin{table}[tb]
\centering
\caption{Classification accuracy at $\tau=0.5$, averaged over the evaluated geometries.}
\label{tab:classification}
\begin{tabular}{lccc}
\hline
Estimator & FNR [\%] & FPR [\%] & F1 \\
\hline
Empirical & 22 & 6 & 0.80 \\
MLP & 10 & \textbf{2} & 0.92 \\
GP & \textbf{6} & 2 & \textbf{0.95} \\
\hline
\end{tabular}
\end{table}

\subsection{LoS Termination Prediction}

The learned skyline is then evaluated as an advance-warning signal for future LoS disappearance events. For this purpose, EphemerisWindow combines the learned blockage probability surface with the future satellite trajectory and predicts whether the satellite will cross the local obstruction boundary within the prediction horizon. The evaluation does not measure the execution of a complete handover procedure. It measures whether the disappearance of the serving or candidate satellite from LoS can be anticipated before the link is lost.

Table~\ref{tab:ephemeriswindow} reports the EphemerisWindow results at $\tau_{\mathrm{pred}}=0.25$. The trajectory-level prediction produces no false alarms in the evaluated cases. Lead time is averaged over detected termination events. The achieved warning depends on the position of the learned $\tau_{\mathrm{pred}}$ contour relative to the true skyline and therefore varies across estimators and geometries.

\begin{table}[tb]
\centering
\caption{LoS termination prediction with EphemerisWindow at $\tau_{\mathrm{pred}}=0.25$.}
\label{tab:ephemeriswindow}
\begin{tabular}{llccc}
\hline
Geometry & Estimator & Detected & Lead [min] & FA [\%] \\
\hline
Canyon & GP & 413 & \textbf{3.2} & \textbf{0} \\
Canyon & MLP & 413 & 0.5 & \textbf{0} \\
Street & GP & 323 & \textbf{44.4} & \textbf{0} \\
Street & MLP & 321 & 12.9 & \textbf{0} \\
\hline
\end{tabular}
\end{table}

The evaluation shows that the GP and MLP estimators address different operating regimes. The GP is preferable for small training data sets, smooth obstruction profiles, and cases where calibrated uncertainty is required. Its main limitation is the cubic training complexity, which makes it less suitable for long-term buffers. The MLP is preferable when the training data set grows, when the obstruction profile becomes more irregular, or when accelerated training is available. The empirical baseline remains useful as a transparent low-complexity reference, but its lack of cross-azimuth generalisation limits its accuracy in sparse observation regions.

\iftoggle{submission}{

The evaluation underlines that the main value of the learned skyline is the separation between geometric satellite visibility and local LoS availability. The ephemeris determines the future trajectory of the satellite, while the learned skyline defines the local horizon line at which the reference signal disappears. By intersecting both, the terminal can estimate the remaining LoS time of the serving satellite and provide the timing input required for handover preparation. Therefore, the proposed method should be interpreted as an advance-warning mechanism for LoS termination, rather than as a complete handover execution procedure.
\section{Conclusion}\label{sec:conclusion}
}{
\subsection{Discussion}

The evaluation underlines that the main value of the learned skyline is the separation between geometric satellite visibility and local LoS availability. The ephemeris determines the future trajectory of the satellite, while the learned skyline defines the local horizon line at which the reference signal disappears. By intersecting both, the terminal can estimate the remaining LoS time of the serving satellite and provide the timing input required for handover preparation. Therefore, the proposed method should be interpreted as an advance-warning mechanism for LoS termination, rather than as a complete handover execution procedure.

The circular azimuth encoding is an important part of this functionality. A scalar azimuth representation introduces an artificial discontinuity between $0^\circ$ and $360^\circ$, although both directions are adjacent in the local sky. This would force the estimator to represent one physical obstruction as two disconnected regions when the obstruction crosses the North direction. The $(\sin\phi,\cos\phi)$ encoding removes this discontinuity and allows the GP kernel and the neural estimator to learn a continuous blockage surface over the full azimuth range.

The threshold $\tau$ provides the operational safety margin of the learned surface. The central contour $\tau=0.5$ represents the most likely obstruction boundary, while lower thresholds define conservative contours that are crossed earlier by the satellite trajectory. This is useful for handover preparation, since the preparation process must start before the current LoS path disappears. A single trained surface can therefore support different operating points by changing $\tau$, without retraining the estimator.

The proposed method also differs from map-based visibility prediction, camera-based skymask estimation, and statistical LoS models. Map-based methods require an environmental model before visibility can be predicted. Camera-based methods require additional optical sensing. Statistical models provide average blockage behaviour over an environment class. In contrast, the proposed method learns the terminal-specific obstruction realisation from passive satellite observations and makes this learned local model queryable along future LEO trajectories.

The main limitations follow from this passive and local formulation.  The chosen constellation also affects the learning process. A denser constellation increases the number of observable satellite directions per time interval and can therefore reduce the time required to populate the training data set. A sparser constellation, or a constellation with an inclination that provides uneven angular coverage at the terminal latitude, may leave azimuth regions weakly observed for longer periods. In these regions, the GP relies on kernel-based interpolation and the neural estimator relies on its smooth inductive bias, which increases the importance of uncertainty reporting. As such, the reported results should be interpreted for the evaluated constellation geometry, while the proposed formulation remains applicable to other LEO deployments as long as their satellite passes provide sufficient angular diversity.

The chosen constellation significantly impacts the learning process. A denser constellation increases the number of observable satellite directions per time interval and can therefore reduce the time required to populate the training data set. A sparser constellation, or a constellation with an inclination that provides uneven angular coverage at the terminal latitude, may leave azimuth regions weakly observed for longer periods. In these regions, the GP relies on kernel-based interpolation and the neural estimator relies on its smooth inductive bias, which increases the importance of uncertainty reporting. As such, the reported results should be interpreted for the evaluated constellation geometry, while the proposed formulation remains applicable to other LEO deployments as long as their satellite passes provide sufficient angular diversity.

Finally, temporary obstructions, vegetation, diffraction, and partial attenuation can introduce inconsistent binary observations near the true boundary, requiring periodic retraining and uncertainty-aware use of the learned skyline.
\section{Conclusions and Future Work}\label{sec:conclusion}
}

This paper has presented a map-free method for learning the local LoS skyline from passive LEO satellite observations. The method uses satellite ephemeris information to determine the expected azimuth and elevation of each satellite and combines this information with binary availability labels derived from periodic radio measurements. The resulting training data set is used to learn a terminal-specific obstruction boundary without relying on 3D building maps, sky cameras, or additional environmental sensors.

The results show that both the Gaussian Process classifier and the neural MLP can reconstruct the LoS skyline more accurately than empirical bracketing. The GP provides the strongest performance for limited training data sets and smoother obstruction profiles, while the neural estimator provides a scalable alternative for larger buffers and more irregular profiles. When the learned skyline is combined with the future satellite trajectory through EphemerisWindow, the terminal can estimate the remaining LoS time of the serving satellite and provide an advance-warning signal for handover preparation. As such, the proposed method does not replace the handover decision procedure, but provides the local visibility information required to prepare it before the current link disappears.

\iftoggle{submission}{}{
Future work will address mobile terminals, where the obstruction profile changes with the terminal position and the training data set must be spatially filtered. Furthermore, sparse Gaussian Process approximations and online neural updates should be investigated to support longer observation windows and continuous operation. Additional work is also required to include partial attenuation, diffraction, vegetation, and temporary obstructions, since these effects turn the binary LoS boundary into a more gradual and time-varying visibility condition. Finally, the integration of the learned skyline with full NTN handover control should be evaluated, including target selection, preparation latency, and make-before-break execution.
}
\section*{Acknowledgment}
The work was co-funded by the German Federal Ministry for Research, Technology and Space under the Open6GHub+ (Grant 16KIS2403) and 6G Coverage (Grant 16KIS2420) projects.
The views and opinions expressed are solely those of the authors and do not represent the official position of the funding authorities.

\bibliographystyle{IEEEtran}
\bibliography{references}

\end{document}